\documentclass{desyproc}

\begin{document}
\title{Latest results of MEG and status of MEG-II}

\author{{\slshape Francesco Renga$^1$ [for the MEG Collaboration]}\\[1ex]
$^1$INFN - Sez. di Roma, P.le A.~Moro 2, 00185 Roma, Italy}

\contribID{xy}

\confID{8648}  
\desyproc{DESY-PROC-2014-04}
\acronym{PANIC14} 
\doi  

\maketitle

\begin{abstract}
Within the Standard Model, in spite of neutrino oscillations, the flavor of charged leptons is conserved in very good approximation, and therefore charged Lepton Flavor Violation 
is expected to be unobservable. On the other hand, most new physics models predict charged Lepton Flavor Violation within the experimental reach, and processes 
like the $\mu \to e \gamma$ decay became standard probes for physics beyond the Standard model. The MEG experiment, at the Paul Scherrer Institute 
(Switzerland), searches for the $\mu \to e \gamma$ decay, down to a Branching Ratio of about $5 \times 10^{-13}$, exploiting the most intense continuous muon beam in the 
world and innovative detectors. In this talk I will present the latest results from MEG, and the status of its upgrade (MEG-II), aiming at an improvement of the sensitivity by 
one order of magnitude within this decade.\end{abstract}

\section{Introduction}
Charged lepton flavor conservation is an accidental symmetry in the standard model (SM), not related to the gauge structure of the theory, but following 
from the particle content of the model. As a consequence, this conservation is naturally violated in most of the extensions of the standard model. Indeed, LFV in the 
charged lepton sector (cLFV) is expected in the SM due to neutrino oscillations, but the expected branching ratios for LFV decays ($< 10^{-40}$) are predicted to be well 
below the current experimental sensitivities. Hence, an observation of cLFV would be an unambiguous evidence of new physics (NP) beyond the SM.

Among the NP models predicting cLFV at observable levels, Supersymmetry (SUSY) is of particular interest: even if the theory is developed to be flavor blind at the high
energy scale, cLFV arises at the electroweak scale through renormalization group equations, and hence it is essentially unavoidable. Moreover, many SUSY models predict
a strong correlation between cLFV and the possible deviation of the muon $g-2$ from its SM prediction. Anyway, the expected branching ratios
strongly depend on the specific flavor structure of the model. Recent limits on $\mu \to e \gamma$ already rule out several scenarios still allowed by direct 
searches at LHC but nonetheless, even within the same models, a different flavor structure can predict rates not yet explored, and within the reach of
the next generation of cLFV experiments (see~\cite{calibbi:2014} for a specific model with flavored gauge mediation).

I will report here the latest results for the search of $\mu \to e \gamma$ with the MEG experiment, and the status of its upgrade MEG-II.

\section{The quest for $\mu^+ \to e^+ \gamma$ with the MEG experiment}

The MEG experiment~\cite{meg_detector}, at the Paul Scherrer Institut (PSI, Switzerland), exploits the most intense continuous muon beam in the 
world (up to $10^8$ muons per second) to search for the $\mu^+ \to e^+ \gamma$ decay. Positive muons are stopped in a thin plastic target, 
and hence the signature of the $\mu^+ \to e^+ \gamma$ decay is given 
by a positron and a photon, monochromatic ($\sim$~52.8~MeV), emitted at the same time, and back-to-back. Although a prompt background is 
given by the radiative $\mu \to e \nu \overline \nu \gamma$ decay, the largely dominant background, when operating with with very high muon beam intensity,
is given by the accidental coincidence of a positron from a muon decay with a photon from another muon decay (radiative decay or annihilation in flight of the positron).
The background rate is then proportional to the square of the muon rate, making useless a further increase of the muon rate as soon as the background expected in the signal region 
becomes relevant. For this reason, the MEG experiment is operated with $\sim 3 \times 10^7$ muons per second, which is found to be an optimal value for our setup.

\begin{figure}[hb]
\vspace{-5.5cm}
\centerline{\includegraphics[width=10cm]{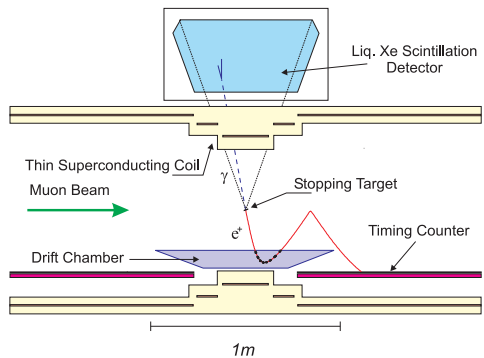}}
\vspace{-3.cm}
\caption{The MEG detector.}\label{fig:detector}
\end{figure}

The MEG detector is shown in Figure~\ref{fig:detector}. Positron are reconstructed in MEG by a system of 16 planar drift chamber in a gradient magnetic field, with its main 
component along the beam axis, and a system of 30 scintillating bars for timing and trigger. The gradient magnetic field is necessary to prevent tracks emitted at almost 90 degrees 
with respect to the beam axis to make several turns within the spectrometer before exiting the detector. The drift chambers reached a resolution of $\sim$~300~$\mu$m in the radial 
direction and $\sim$~1~mm along the beam axis, resulting into a core momentum resolution of $\sim$~330~keV and angular resolutions of $\sim$~10~mrad. The timing counter 
allows to measure the positron time with a resolution of $\sim$ 70~ps. The overall positron efficiency is $\sim$~30\%, and it is largely dominated by the loss of positrons in the path 
from the drift chamber system to the timing counter.

Photons are reconstructed by a liquid Xenon detector instrumented with 856 PMTs. It measures the energy, the time and the conversion point of the photon, with resolutions of 
$\sim$~900~keV in the bulk region of the detector, 70~ps and $\lesssim$~6~mm. 

The decay vertex is defined by the intersection of the positron track with the target, while the direction from the vertex to the photon conversion point is taken as the photon direction
to determine the relative $e \gamma$ angle.

Electronic waveforms from all detectors are fully digitized at GS/s rates thanks to the DRS4 chip developed at PSI. A fully digital trigger system has been developed,
exploiting energy, time and position measurements in the Xenon detector and time measurement in the timing counter.

Several calibrations are necessary to reach and measure the mentioned resolutions. Among them, it is worth to mention the use of a pion beam, along with an ancillary 
photon detector to select back to back photon pairs in the reaction chain $\pi^- + p \to \pi^0 + n$, $\pi^0 \to \gamma \gamma$. Kinematics make the selected positrons 
almost monochromatic, with an energy of about 55~MeV, very near to the signal photon energy. This is used to calibrate the absolute energy scale of the calorimeter, which is 
then monitored periodically with low energy photons from proton-induced nuclear reactions, in order to finally get a 0.2\% accuracy on the energy scale.

\begin{figure}[hb]
\centerline{\includegraphics[width=10cm]{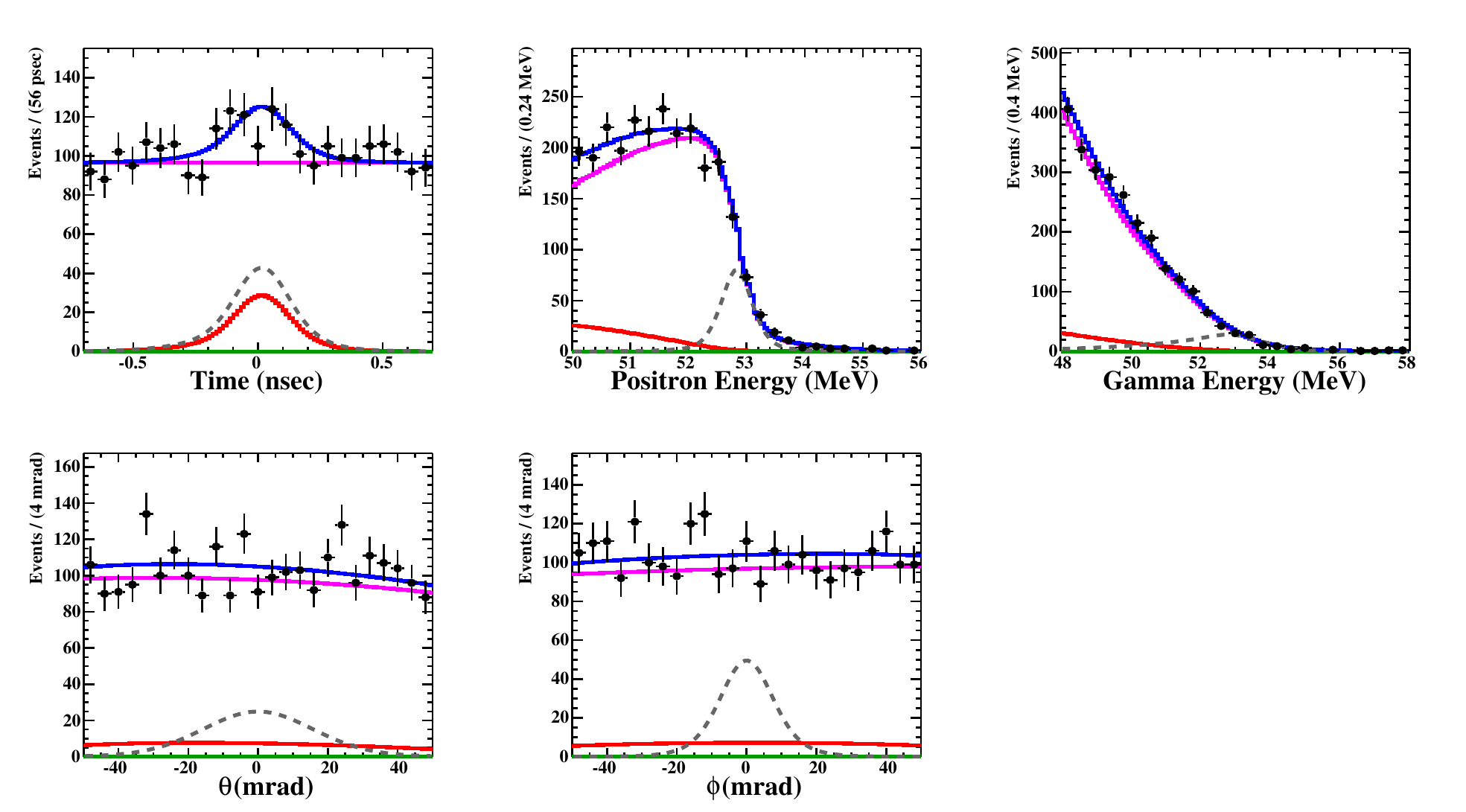}}
\caption{Result of the fit (blue) to the 2009-2011 data (black dots with error bars). Contributions from accidental background (blue), radiative decay background (red) and 
signal (green) are shown. The signal PDFs for a branching ratio of $3 \times 10^{-11}$ are also shown for reference (gray).}\label{fig:result_2013}
\end{figure}

A likelihood analysis is used for the search of $\mu \to e \gamma$. Five discriminating variables are used: the positron energy $E_e$, 
the photon energy $E_\gamma$, the relative time $T_{e\gamma}$ and the projections of the relative angle angle $\phi_{e\gamma}$ and $\theta_{e\gamma}$. The signal Probability Distribution
Functions (PDFs) are obtained by combining the measured resolution, as well as the PDFs for the radiative muon decays. Conversely, the PDFs for the accidental background are fully
extracted from data, using sideband regions defined in the $E_\gamma$ vs. $T_{e\gamma}$ plane. In the construction of the PDFs we take into account several correlations in the positron 
PDFs, which emerge from geometrical effects and are well understood both qualitatively and quantitatively. The results of the likelihood fit, based on the data collected in the 2009-2011
period ($\sim 36 \times 10^{13}$ muons stopped in the target), are shown in Figure ~\ref{fig:result_2013}~\cite{meg_2013}. No significant signal is observed, and an upper limit 
on the number of signal events has been extracted at 90\% confidence level, with a frequentistic approach based on a profile likelihood ratio. This is combined with a
normalization factor obtained by counting the number of Michel positrons reconstructed in the spectrometer and, including all systematics (dominated by the uncertainties on the 
PDFs for the relative angle), an upper limit of $5.7 \times 10^{-13}$ is obtained for the $\mu \to e \gamma$ branching ratio, to be compared with an expected limit of 
$7.7 \times 10^{-13}$ (from toy Monte Carlo studies).

The MEG experiment collected data in 2012 and 2013, and doubled the available statistics. Several improvements have been included in the on going analysis of these data: 
refined algorithms allowed to increase the efficiency for tracks making several turns within the spectrometer; a more accurate measurement if the magnetic field has been 
performed in order to reduce the corresponding systematic uncertainties; an algorithm for the recognition of photons coming from positron annihilation in flight has been 
introduced, in order to suppress the main contribution to the background at large photon energies. Although the determination of the final sensitivity is still ongoing, 
an expected upper limit below $5 \times 10^{-13}$ is foreseen.
 
\section{Status of MEG-II}

As already mentioned, many NP models predict a $\mu \to e \gamma$ branching ratio not far from the current limit. Hence, a short term upgrade to improve the 
sensitivity of MEG of about one order of magnitude is worth the effort. As shown in~\cite{meg_upgrade}, the upgraded experiment will be competitive with the first 
phase of the experiments searching for $\mu \to e$ conversion in the field of nuclei, if cLFV arises from magnetic-moment operators, 
$\mathcal{L} \propto \overline \mu_R \sigma_{\mu\nu}e_LF^{\mu\nu}$, like in supersymmetry.

The upgrade of the MEG experiment will involve all the subdetectors. The system of 16 drift chambers will be replaced with a unique cylindrical drift chamber with stereo wires.
The chamber will be operated with a light mixture of Helium and Isobutane (85\%:15\%) to reduce the material budget, and will cover all the path of the track to the timing counter, in 
order to recover the large inefficiency observed in MEG. A single hit resolution of about 120 $\mu$m is expected and confirmed by measurements with different prototypes in different 
environments (cosmic rays and positron beams). Given the stereo angle, it will give a resolution below 1 mm along the beam axis. A momentum resolution of 130 keV and angular 
resolutions  of about 5~mrad are expected.

The Timing Counter will be replaced by about 500 scintillating tiles read out by Silicon Photomultipliers (SiPM). Tests with positron beams confirmed that a time resolution of about 30 ps
can be reached combining the time measurements of the tiles hit by each track.

The PMTs in the inner face of the liquid Xenon calorimeter will be replaced with SiPMs, specifically developed to be sensitive to ultra-violet scintillation light of Xenon.
The improved granularity will allow to improve the resolutions for photons converting just after entering the calorimeter (\emph{shallow events}) and the capability of detecting 
pileup photons. Moreover, the geometry of the lateral faces will be changed in order to increase the fiducial volume and better control the reflection of the scintillation light. 
The energy resolution is expected to go down to 1\%. Finally, a new DAQ board is under development, still based on the DRS4 chip, in order to handle the increased number 
of channels within the limited space of the MEG experimental hall. 

The upgraded detectors are presently under construction and are expected to be ready for an engineering run at the end of 2015 and for physics runs in 2016. A three-year data taking 
campaign is foreseen, with an optimal muon rate which should reach $7 \times 10^{7}$ muons per second, thanks to the improved resolutions. An expected upper limit 
of about $4 \times 10^{-14}$ is finally envisaged.


\begin{footnotesize}



%

\end{footnotesize}


\end{document}